%% file: paper.tex
\documentclass[usenatbib,useAMS]{mn2e}

\usepackage{times}
\usepackage{rotating}
\usepackage{graphicx}
\usepackage{epstopdf}
\usepackage{amsmath}
\usepackage{amssymb}
\usepackage{bbold}
\usepackage{multirow}
\usepackage{url}

\usepackage[rightcaption]{sidecap}

\usepackage{xcolor}

\usepackage{float}

\input{./HEADER}

\begin{document}

\title[Perturbation Theory and Stochastic Halo Biasing]{{Modelling Baryon Acoustic Oscillations with Perturbation Theory and Stochastic Halo Biasing}}

\author[F.~S.~Kitaura, G.~Yepes \& F.~Prada]{Francisco-Shu Kitaura$^{1}$\thanks{E-mail: kitaura@aip.de, Karl-Schwarzschild-fellow}, Gustavo Yepes$^{2}$ \& Francisco Prada$^{3,4,5}$ \\
$^{1}$Leibniz-Institut f\"ur Astrophysik Potsdam (AIP), An der Sternwarte 16, D-14482 Potsdam, Germany\\
$^{2}$Departamento de F{\'i}sica Te{\'o}rica,  Universidad Aut{\'o}noma de Madrid, Cantoblanco, 28049, Madrid, Spain\\
$^{3}$Campus of International Excellence UAM+CSIC, Cantoblanco, E-28049 Madrid, Spain \\
$^{4}$Instituto de F{\'i}sica Te{\'o}rica, (UAM/CSIC), Universidad Aut{\'o}noma de Madrid, Cantoblanco, E-28049 Madrid, Spain\\
$^{5}$Instituto de Astrof{\'i}sica de Andaluc{\'i}a (CSIC), Glorieta de la Astronom{\'i}a, E-18080 Granada, Spain\\
}

\maketitle

\input{./abstract}

\input{./main}

{\small
\bibliographystyle{mn2e}
\bibliography{lit}
}

\end{document}

%% file: HEADER.tex
\newcommand{\mbi}[1]{\mbox{\boldmath$#1$}}

\newcommand{\lsim}{\mbox{${\,\hbox{\hbox{$ < $}\kern -0.8em \lower 1.0ex\hbox{$\sim$}}\,}$}}
\newcommand{\gsim}{\mbox{${\,\hbox{\hbox{$ > $}\kern -0.8em \lower 1.0ex\hbox{$\sim$}}\,}$}}

\def\beqn{\vspace{2mm}
\begin{eqnarray}} 
\def\eeqn{\vspace{2mm} 
\end{eqnarray}}

 \def\la{\langle}

\newcommand{\be}{\begin{equation}}
\newcommand{\ee}{\end{equation}}
\newcommand{\ba}{\begin{eqnarray}}
\newcommand{\ea}{\end{eqnarray}}
\newcommand{\brr}{\begin{array}}
 
\newcommand{\err}{\end{array}}
\newcommand{\bc}{\begin{center}}
\newcommand{\ec}{\end{center}}

%% file: abstract.tex
\begin{abstract}

In this work we investigate the generation of mock halo catalogues based on perturbation theory and nonlinear stochastic biasing with the novel \textsc{patchy}-code. In particular, we use Augmented Lagrangian Perturbation Theory (ALPT) to generate a dark matter density field on a mesh starting from Gaussian fluctuations and to compute the peculiar velocity field. ALPT is based on a combination of second order LPT (2LPT) on large scales and the spherical collapse model on smaller scales. We account for the systematic deviation of perturbative approaches from $N$-body simulations together with halo biasing adopting an exponential bias model. We then account for stochastic biasing by defining three regimes: a low, an intermediate and a high density regime, using a Poisson distribution in the intermediate regime and the negative binomial distribution --including an additional parameter-- to model over-dispersion in the high density regime. Since we focus in this study on massive halos, we suppress the generation of halos in the low density regime. The various nonlinear and stochastic biasing parameters, and density thresholds are calibrated with the large BigMultiDark $N$-body simulation to match the power spectrum of the corresponding halo population. Our model effectively includes only five parameters, as they are additionally constrained by the halo number density. Our mock catalogues show power spectra, both in real- and redshift-space,  which are compatible with $N$-body simulations within about 2\% up to $k\sim 1\, h$ Mpc$^{-1}$ at $z=0.577$ for a sample of halos with the typical BOSS CMASS galaxy number density. The corresponding correlation functions are compatible down to a few Mpc. We also find that neglecting over-dispersion in high density regions produces power spectra with deviations of 10\% at $k\sim 0.4\, h$ Mpc$^{-1}$. These results indicate the need to account for an accurate statistical description of the galaxy clustering for precise studies of large-scale surveys.

\end{abstract}


%% file: main.tex
\section{Introduction}

The new generation of galaxy surveys
request precise numerical simulations of structure formation to compare theoretical models to observations. This is computationally very demanding as the parameter space one needs to cover is extremely large, ranging from varying the cosmological parameters, over modelling different biased tracers, to account for cosmic variance \citep[for large-volume $N$-body simulations see e.g.][]{KimHorizon2009,Prada2012,AnguloXXL2012,DeusSimulation2012,jubilee2013}.
As an alternative to run $N$-body cosmological simulations for each parameter set, one can calibrate approximate structure formation models to $N$-body solutions and scan the parameter space using the more efficient schemes.
  A number of approaches has been proposed in the literature for the generation of mock galaxy catalogues based on Lagrangian Perturbation Theory (LPT), such as \textsc{Pinocchio} \citep[][]{2002ApJ...564....8M,monaco2013} or \textsc{PThalos} \citep[][]{scocci,manera12}.  It has been shown  that perturbation theory can provide an accurate  approach to model Baryon Acoustic Oscillations (BAOs) \citep[][]{2012JCAP...04..013T}. 
The uncertainty of a few Mpc in the position of dark matter particles (or halos) following the approximate schemes is translated into a damping of the power spectrum, which may be modeled by a Gaussian smoothing of the typical uncertainty scale \citep[][]{monaco2013}.
As a consequence, the power spectra predicted by perturbation theory lie below the linear power spectrum instead of developing the characteristic nonlinear excess of power with respect to the linear power spectrum at modes $k\gsim 0.1 \,h$ Mpc$^{-1}$.
Interesting alternatives have been recently proposed, such as re-scaling $N$-body simulations to account for a change in the cosmological parameters  \citep[][]{2010MNRAS.405..143A}, compute covariance matrices from a set of small-volume  simulations \citep[][]{2011ApJ...737...11S}, or including 2LPT within the Vlasov equations solver to speed up  $N$-body codes \citep[\textsc{Cola},][]{cola2013}. 
In this letter, we propose to use an extremely efficient approach based on low resolution one-step perturbation theory solvers.
We rely on Augmented LPT (ALPT), which is based on a combination of second order LPT on large scales with the spherical collapse model on smaller scales, suppressing in this way shell-crossing with an improved modelling of filaments  \citep[][]{2012arXiv1212.3514K}. In this work, we introduce the peculiar velocity within this formalism to model redshift-space distortions. 

To account for the missing power of perturbative approaches at high modes, and at the same time for the scale-dependent bias of halos, we use an exponential bias \citep[][]{Cen1993}. Such a model has been recently proposed to sample halos below the resolution of dark matter simulations \citep[][]{delaTorre2012}. This model is related to the lognormal model \citep[][]{1991MNRAS.248....1C}, and thus to the linear component of the density field \citep[][]{kitlin}, solving the negative densities problem \citep[][]{kitaura_log} of \citet[][]{1993ApJ...413..447F}'s  formulation. Here, we propose to model the statistics of halos with a Poissonian and a negative  binomial distribution function depending on the density regime. The required parameters in our model are calibrated with one of the new set of  the publicly available {BigMultiDark} simulations\footnote{http://www.multidark.org}  \citep[][]{hessBMD}. 

Our approach is not only useful to generate mock catalogues, but also for inference analysis of the large-scale structure (density fields, power spectra, etc), improving previous models based on a linear bias and on the Poisson assumption \citep[see e.g.][]{kitaura,kitaura_log}.

This letter is structured as follows: in the next section (\S \ref{sec:method}) we present our method. We then show  (\S \ref{sec:results}) our numerical experiments calibrating our mock catalogues with $N$-body simulations. Finally (\S \ref{sec:conc}) we present our conclusions and discussion.

\section{Method}

\label{sec:method}

Our approach combines an efficient structure formation model with a local, nonlinear, scale-dependent and stochastic biasing scheme. The resulting computer code is dubbed \textsc{patchy} ({{\bf P}erturb{\bf A}tion {\bf T}heory {\bf C}atalog generator of {\bf H}alo and galax{\bf Y} distributions}).

\subsection{Structure formation model}

We use  Augmented Lagrangian Perturbation Theory (ALPT) to simulate structure formation \citep[][]{2012arXiv1212.3514K}.
In this approximation the displacement field  $\mbi\Psi(\mbi q,z)$, mapping a distribution of dark matter particles at initial Lagrangian positions $q$ to the final Eulerian positions $\mbi x(z)$ at redshift $z$ ($\mbi x(z)=\mbi q+\mbi\Psi(\mbi q,z)$), is split into a long-range $\mbi\Psi_{\rm L}(\mbi q,z)$ and a short-range component $\mbi\Psi_{\rm S}(\mbi q,z)$, i.e. 
$\mbi\Psi(\mbi q,z)=\mbi\Psi_{\rm L}(\mbi q,z)+\mbi\Psi_{\rm S}(\mbi q,z)$.
We rely on 2LPT for the long-range component: 
\be
\mbi\Psi_{\rm 2LPT} =  - D\nabla\phi^{(1)} + D_2\nabla\phi^{(2)}\,,
\ee
where $D$ is the linear growth factor and $D_2 \simeq-3/7\,\Omega^{-1/143}D^2$ \citep[for details on 2LPT see][]{1994MNRAS.267..811B,bouchet1995,catelan}. The potentials $\phi^{(1)}$ and $\phi^{(2)}$ are obtained by solving a pair of Poisson equations: $\nabla^2\phi^{(1)} = \delta^{(1)}$, where $\delta^{(1)}$ is the linear overdensity, and $\nabla^2\phi^{(2)} =  \delta^{(2)}$. The second order nonlinear term $\delta^{(2)}$ is fully determined by the  linear overdensity field $\delta^{(1)}$ through the following quadratic expression: 
\be
\delta^{(2)}\equiv\sum_{i>j} \Big( \phi^{(1)}_{,ii}\phi^{(1)}_{,jj}-[\phi^{(1)}_{,ij}]^2\Big)\,,
\ee 
where we use  the following notation $\phi_{,ij} \equiv \partial^2\phi/\partial q_i\partial q_j$, and the indices $i,j$ run over the three Cartesian coordinates.

 The resulting displacement field is filtered  with a kernel $\cal K$: $\mbi\Psi_{\rm L}(\mbi q,z)={\cal K}(\mbi q,r_{\rm S}) \circ \mbi\Psi_{\rm 2LPT}(\mbi q,z)$.
We apply a Gaussian filter ${\cal K}(\mbi q,r_{\rm S})$$=\exp{(-|\mbi q|^2/(2r_{\rm S}^2))}$, with $r_{\rm S}$ being the smoothing radius.
 We use the  spherical collapse approximation to model the short-range component $\mbi\Psi_{\rm SC}(\mbi q,z)$ \citep[see][]{1994ApJ...427...51B,2006MNRAS.365..939M,2013MNRAS.428..141N}:  $\mbi\Psi_{\rm S}(\mbi q,z)=\left(1-{\cal K}(\mbi q,r_{\rm S}) \right)\circ \mbi\Psi_{\rm SC}(\mbi q,z)$, where 
\be
\label{eq:sc} 
 \mbi \Psi_{\rm SC}  = \nabla\nabla^{-2} \, \left[ 3 \left( \left(1-\frac{2}{3}D\delta^{(1)}\right)^{1/2}-1\right)\right]\,.
\ee
The combined ALPT displacement field
\be
\label{eq:disp}
\mbi\Psi_{\rm ALPT}(\mbi q,z)={\cal K}(\mbi q,r_{\rm S}) \circ \mbi\Psi_{\rm 2LPT}(\mbi q,z)+\left(1-{\cal K}(\mbi q,r_{\rm S}) \right)\circ \mbi\Psi_{\rm SC}(\mbi q,z)
\ee
 is used to move a set of homogenously distributed particles from Lagrangian initial conditions to the Eulerian final ones. We then grid the particles following a clouds-in-cell scheme to produce a smooth density field $\delta^{\rm ALPT}$.

\begin{figure}
\vspace{-1cm}
\begin{tabular}{cc}
\hspace{3.8cm}
\includegraphics[width=4.5cm]{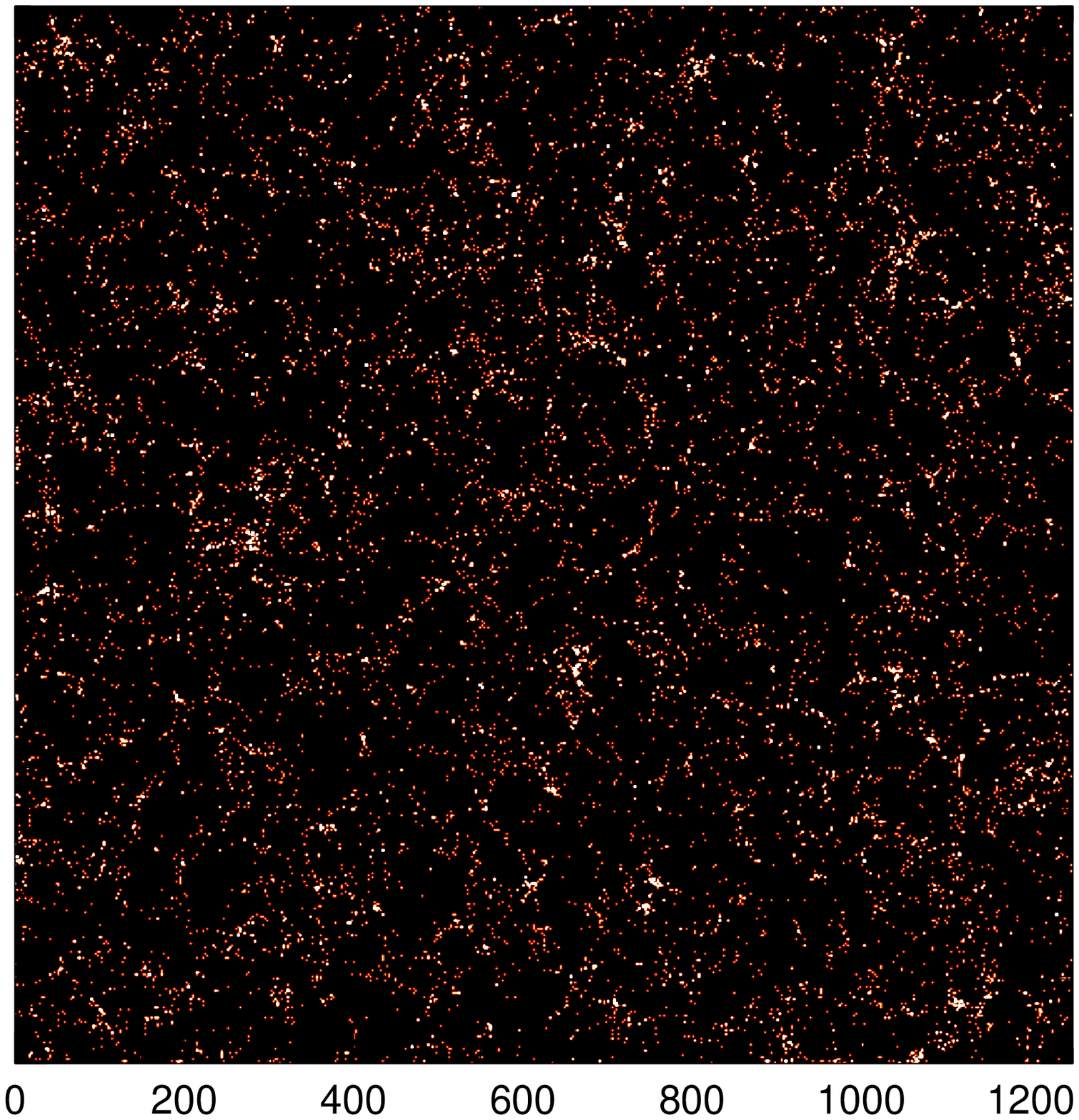}
\hspace{-8.6cm}\includegraphics[width=4.5cm]{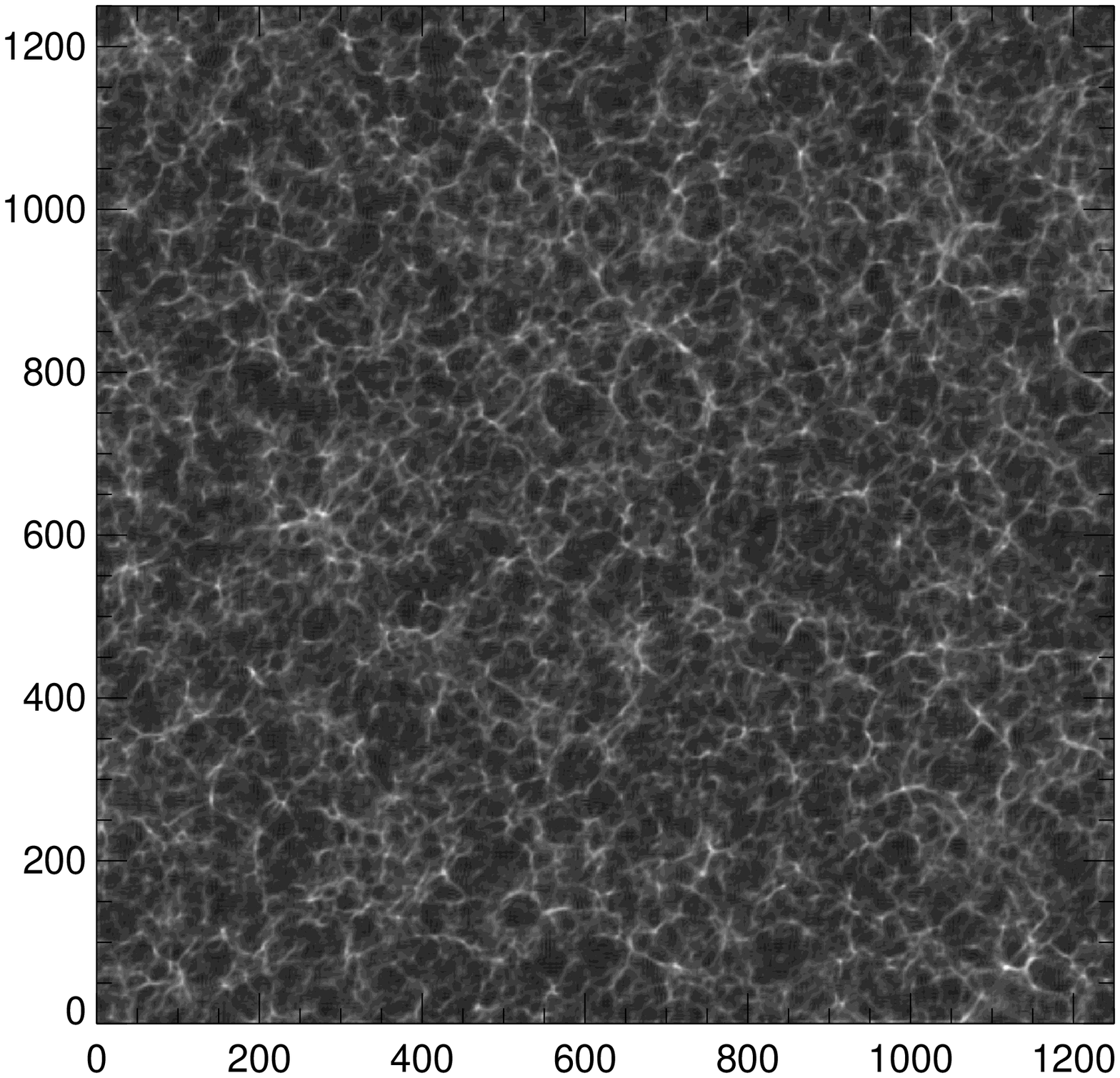}
\put(-135,85){\rotatebox[]{90}{$Y$ [$h^{-1}$ Mpc]}}
\put(-80,11){$X$ [$h^{-1}$ Mpc]}
\put(35,11){$X$ [$h^{-1}$ Mpc]}
\end{tabular}
\vspace{-0.5cm}
\caption{
\label{fig:2dplots} 
Slices of thickness 20 $h^{-1}$ Mpc and  1250 $h^{-1}$ Mpc side  of a \textsc{patchy} simulation through the dark matter density field (on the left) and through the corresponding halo field (on the right). The logarithm of the density fields are shown. Lighter regions represent higher densities.}
\end{figure}


\begin{figure*}
\begin{tabular}{cc}
\includegraphics[width=7.cm]{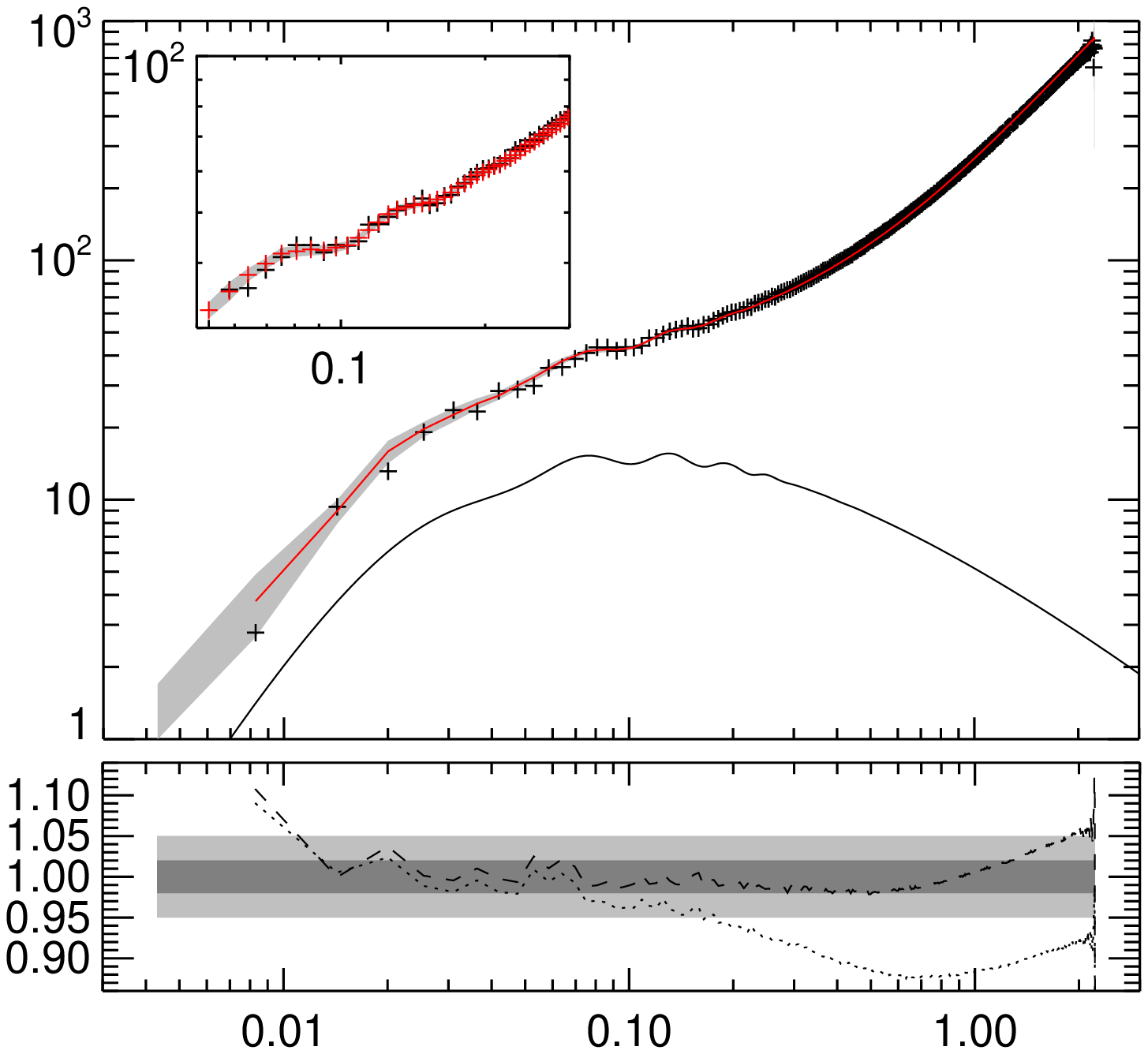}
\put(-100,72){real-space}
\hspace{-1.2cm}
\includegraphics[width=7.cm]{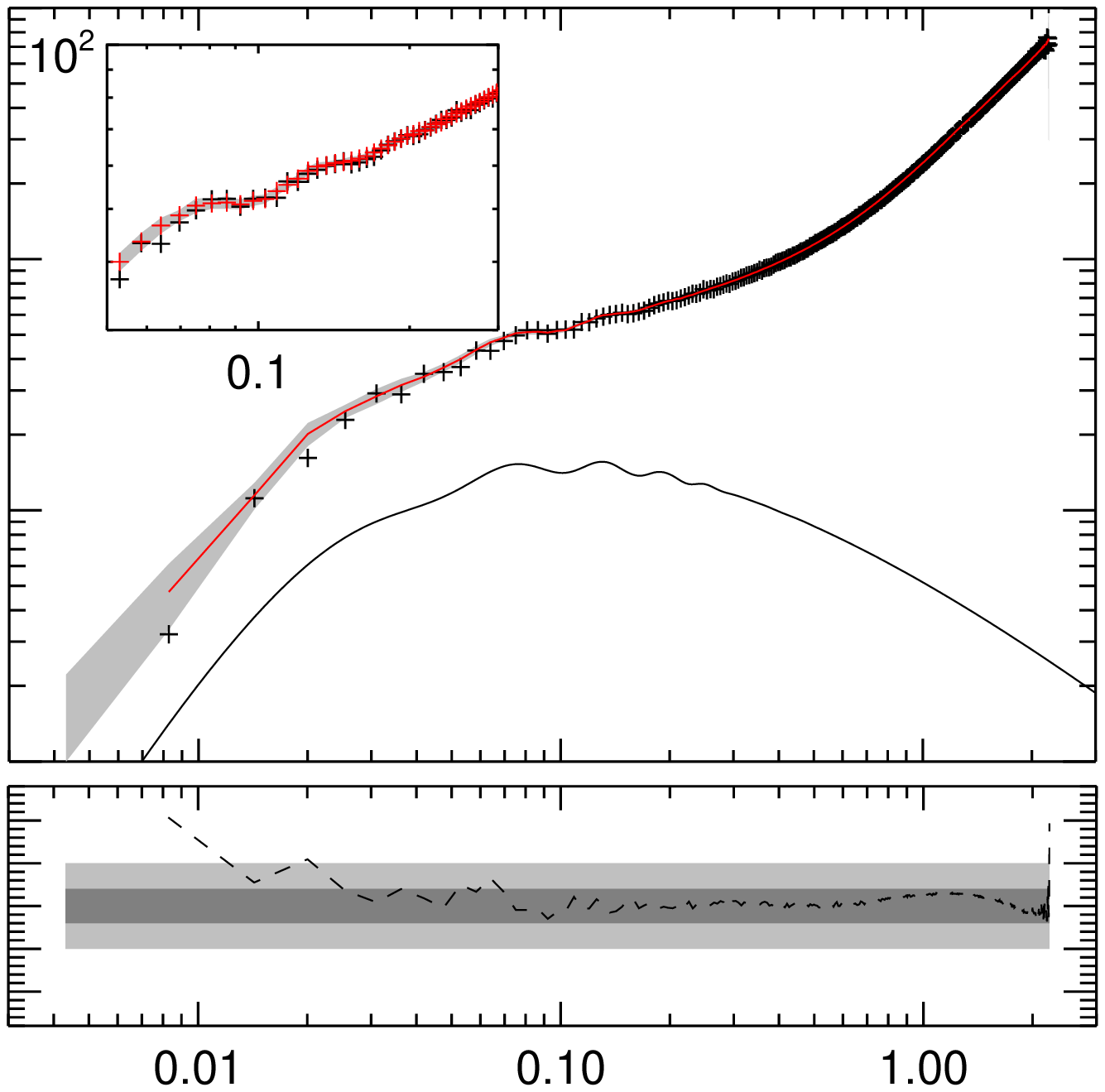}
\put(-100,72){redshift-space}
\put(-365,120){\rotatebox[]{90}{$P(k)\times k^{1.5}$}}
\put(-380,30){\rotatebox[]{90}{$\langle P^{\rm Patchy}(k)\rangle/$}}
\put(-370,28){\rotatebox[]{90}{$\langle P^{{\rm Nbody}}(k)\rangle$}}
\put(-110,-5){$k$ [$h$ Mpc$^{-1}$]}
\put(-280,-5){$k$ [$h$ Mpc$^{-1}$]}
\end{tabular}
\caption{\label{fig:ps} Power spectra obtained with \textsc{patchy} vs. {BigMultiDark} at $z=0.577$ for a halo sample with number density 3.6 $\times$ 10$^{-4}$ Mpc$^{-3}$ $h^3$ in real-pace (on the left) and in redshift-space (on the right). The red line corresponds to the mean of 50 \textsc{patchy} realizations with the corresponding 1-sigma region in grey. The linear power spectrum is also shown (solid black line) as well as the mean over 8 sub-volumes of the {BigMultiDark} simulation. Bottom: Ratio between the mean of the \textsc{patchy} realizations and the mean of the $N$-body sub-volumes. Dotted curve: only Poisson, dashed curves: includes negative-binomial PDF. Regions within 2\% are inidcated by the dark grey area and 5\% by the lighter one.}
\end{figure*}

\subsection{Deterministic biasing}

The relation between the halo distribution and the underlying dark matter density field is known to be nonlinear, nonlocal and stochastic \citep[][]{1974ApJ...187..425P,1985MNRAS.217..805P,1986ApJ...304...15B,1993ApJ...413..447F,1996MNRAS.282..347M,1999ApJ...520...24D,1999MNRAS.304..767S,2000MNRAS.318..203S,2002ApJ...575..587B,2007PhRvD..75f3512S,2010PhRvD..82j3529D,2011PhRvD..83j3509B,2011A&A...527A..87V,2012MNRAS.421.3472E,2012PhRvD..85h3509C,2012PhRvD..86h3540B,baldauf2013}.
We will neglect for the time being  nonlocal biasing. To account for nonlinear biasing we consider an exponential expression which includes only 2 parameters (with only one free parameter), $f_N$ and $\alpha$, to get the expected number counts of halos in a cell $i$ from the density field \citep[][]{Cen1993,delaTorre2012}, i.e. 
\be
\lambda_i\equiv\langle N_i\rangle=f_N\times(1+\delta^{\rm ALPT}_i)^{\rm \alpha}\;,
\ee 
where the brackets stand for the ensemble average over the realization of halos given a particular probability distribution function (PDF) $P(N_i\mid\lambda_i,\{p_i\})$ with a set of parameters $\{p_i\}$: $\langle N_i\rangle =\sum_{N_i=0}^{\infty}P(N_i\mid\lambda_i,\{p_i\})\,N_i$ (see next subsection).
The parameter $\alpha$ controls the nonlinear, scale-dependent bias, while $f_N$ controls the halo number density. We note that for the unbiased case ($\alpha=1$) $f_N$ is equal to the number density  $\overline{N}\equiv\langle\lambda\rangle_V=\langle f_N\times(1+\delta^{\rm ALPT})\rangle_V=f_N$, where $\langle\dots\rangle_V$ is the ensemble average over a sufficiently large volume $V$, so that $\langle\delta^{\rm ALPT}\rangle_V=0$.
Accordingly, we find $f_N$  being in general given by: 
\be
f_N={\overline{N}}/{\langle(1+\delta^{\rm ALPT}_i)^{\rm \alpha}\rangle_V}\,. 
\ee

\subsection{Stochastic biasing}

The halo distribution is a discrete sample of the continuous underlying dark matter distribution. To account for the shot noise one could do Poissonian realizations of the halo density field as given by the deterministic bias and the dark matter field \citep[see e.g.][]{delaTorre2012}. However, it is known that the excess probability of finding halos  in high density regions generates over-dispersion \citep[][]{2001MNRAS.320..289S,2002MNRAS.333..730C}, underdense regions are under-dispersed and  there is an intermediate regime (mainly filaments)  in which Poissonity approximately holds.
Therefore we define three regimes: a low density ($\delta^{\rm ALPT}>\delta^{\rm low}$), an intermediate ($\delta^{\rm low}<\delta^{\rm ALPT}<\delta^{\rm high}$) and a high density regime ($\delta^{\rm ALPT}>\delta^{\rm high}$). We use a Poisson distribution in the intermediate regime: 
\be
P(N_i\mid\lambda_i)=\frac{\lambda_i^{N_i}}{N_i!} \exp(-\lambda_i)\;,
\ee 
and the negative binomial (NB) PDF \citep[for non-Poissonian distributions see][]{1984ApJ...276...13S,1995MNRAS.274..213S}  including an additional parameter $\beta$ to model over-dispersion in the high density regime: 
\be
P(N_i\mid\lambda_i,\beta)=\frac{\lambda_i^{N_i}}{N_i!}\frac{\Gamma(\beta+N_i)}{\Gamma(\beta)(\beta+\lambda)^{N_i}}\frac{1}{(1+\lambda/\beta)^\beta} \;.
\ee 
This PDF tends towards the Poisson distribution for $\beta\rightarrow\infty$. The NB is also very close to the Poissonian distribution for low $\lambda$ values. For this reason we could consider only one threshold density $\delta^{\rm th}=\delta^{\rm low}=\delta^{\rm high}$, reducing the number of parameters in our model. We will investigate this further in future works.
 Since we focus in this study on massive halos to model Luminous Red Galaxies (LRGs), we then suppress the generation of halos in the low density regime.
A Bayesian inference algorithm based on a combination of these PDFs will appear in a forthcoming publication.

\subsection{Redshift-space distortions}

The mapping between Eulerian real-space $\mbi x(z)$ and redshift-space $\mbi s(z)$ is given by: $\mbi s(z)=\mbi x(z)+\mbi v_r(z)$, with $\mbi v_r\equiv(\mbi v\cdot\hat{\mbi r})\hat{\mbi r}/(Ha)$; where  $\hat{\mbi r}$ is the unit sight line vector, $H$ the Hubble constant, $a$ the scale factor, and $\mbi v=\mbi v(\mbi x)$ the 3-d velocity field interpolated at the position of each halo in Eulerian-space $\mbi x$ using the displacement field $\mbi\Psi_{\rm ALPT}(\mbi q,z)$.
We split the peculiar velocity field into a coherent $\mbi v^{\rm coh}$ and a (quasi-) virialized component $\mbi v_{\sigma}$: $\mbi v=\mbi v^{\rm coh}+\mbi v^{\sigma}$. 
The coherent peculiar velocity field is computed in Lagrangian-space from the linear Gaussian field $\delta^{(1)}(\mbi q)$ using the ALPT formulation consistently with the displacement field (see Eq.~\ref{eq:disp}):
\be
\mbi v_{\rm ALPT}^{\rm coh}(\mbi q,z)={\cal K}(\mbi q,r_{\rm S}) \circ \mbi v_{\rm 2LPT}(\mbi q,z)+\left(1-{\cal K}(\mbi q,r_{\rm S}) \right)\circ \mbi v_{\rm SC}(\mbi q,z)
\ee
For the second order LPT component ${\mbi v}_{\rm 2LPT}$ we refer to e.~g.~\citet[][]{1993MNRAS.264..375B,bouchet1995}: 
\be
\label{eq:2lptvel}
 {\mbi v}_{\rm 2LPT} =-fHaD\nabla\phi^{(1)}+f_2HaD_2\nabla\phi^{(2)}\,, 
\ee
where $f_i = {\rm d}\ln D_i/{\rm d}\ln a$ ($D\equiv D_1$, $f\equiv f_1\approx\Omega^{5/9}$, $f_2\approx2\Omega^{6/11}$).


 The spherical collapse component is obtained by performing the time derivative of Eq.~\ref{eq:sc}: 
\be
\label{eq:scvel} 
\mbi v_{\rm SC} =  \nabla\nabla^{-2}\,\left[-{fHaD\,\delta^{(1)}}{\left(1-\frac{2}{3}D\delta^{(1)}\right)^{-1/2}}\right]\,.
\ee
We use the high correlation between the local density field and the velocity dispersion to model the displacement due to (quasi-) virialized motions. Effectively, we sample a Gaussian distribution function ($\mathcal G$) with a dispersion given by $\sigma_v\propto\left(1+b^{\rm ALPT}\delta^{\rm ALPT}\left(\mbi x\right)\right)^\gamma$. Consequently, 
\be
\mbi v^{\sigma}_r\equiv(\mbi v^{\sigma}\cdot\hat{\mbi r})\hat{\mbi r}/(Ha)={\mathcal G}\left(g\times\left(1+b^{\rm ALPT}\delta^{\rm ALPT}\left(\mbi x\right)\right)^\gamma\right)\hat{\mbi r}\,,
\ee 
\citep[see][]{Kitaura07,kitaura,hesscs}. The linear bias between the ALPT approximation and the full $N$-body solution is given by $b^{\rm ALPT}$ and is close to unity for scales of a few Mpc \citep[see][]{hesscs}.
 The parameters $g$ and $\gamma$ have been adjusted to fit the damping effect in the power-spectrum in redshift-space as found in the BigMultiDark $N$-body simulation. In closely virialized systems the kinetic energy approximately equals the gravitational energy and a Kepplerian law predicts $\gamma$ close to $0.5$ \citep[see][]{Kitaura07,hessdv}, leaving only the proportionality constant $g$ as a free parameter in our model.

\section{Numerical experiments}
\label{sec:results}

We use a reference halo catalogue at redshift $z=0.577$ extracted  from one of the  {BigMultiDark} simulations \citep[][]{hessBMD}, which was performed using \textsc{gadget-2} \cite[][]{gadget2} with $3840^3$ particles on a volume of $(2500\, h^{-1}$ Mpc)$^3$ assuming $\Lambda$CDM-cosmology with \{$\Omega_{\rm M}=0.29,\Omega_{\rm K}=0,\Omega_\Lambda=0.71,\Omega_{\rm B}=0.047,\sigma_8=0.82,w=-1,n_s=0.95$\} and a Hubble constant ($H_0=100\,h$ km s$^{-1}$ Mpc$^{-1}$) given by  $h=0.7$. Halos were produced based on density peaks including substructures  using the Bound Density Maximum (BDM) halo finder \citep[][]{klypin1997} and then selected according to a maximum circular velocity larger than 350 km s$^{-1}$ to match the number density of BOSS CMASS galaxies \citep[][]{nuza2013}. For the impact of these selection criteria in the clustering and scale-dependent bias see \citet[][]{pradaBMD}.

\begin{figure}
\hspace{4cm}
\begin{minipage}[t]{0.8\textwidth}
\includegraphics[width=4.5cm]{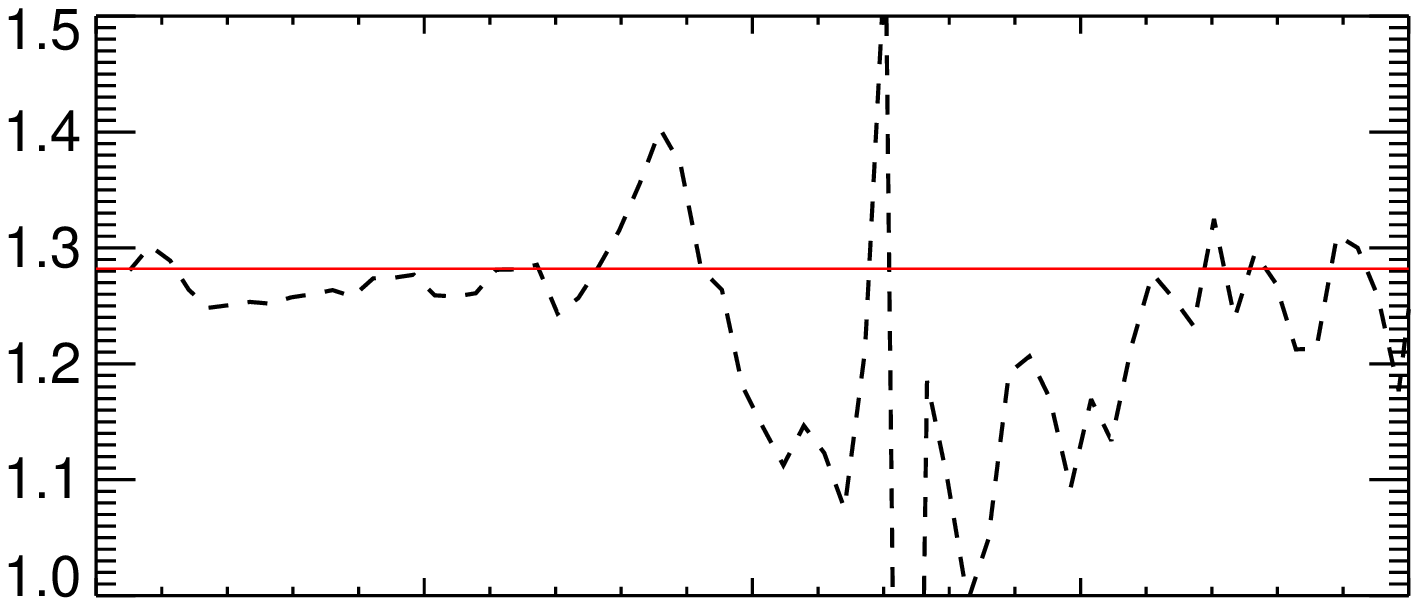}
\put(-132,30){\rotatebox[]{90}{$\xi^z(r)/\xi(r)$}}
\put(-100,20){$N$-body}
\vspace{-.5cm}
\\
\includegraphics[width=4.5cm]{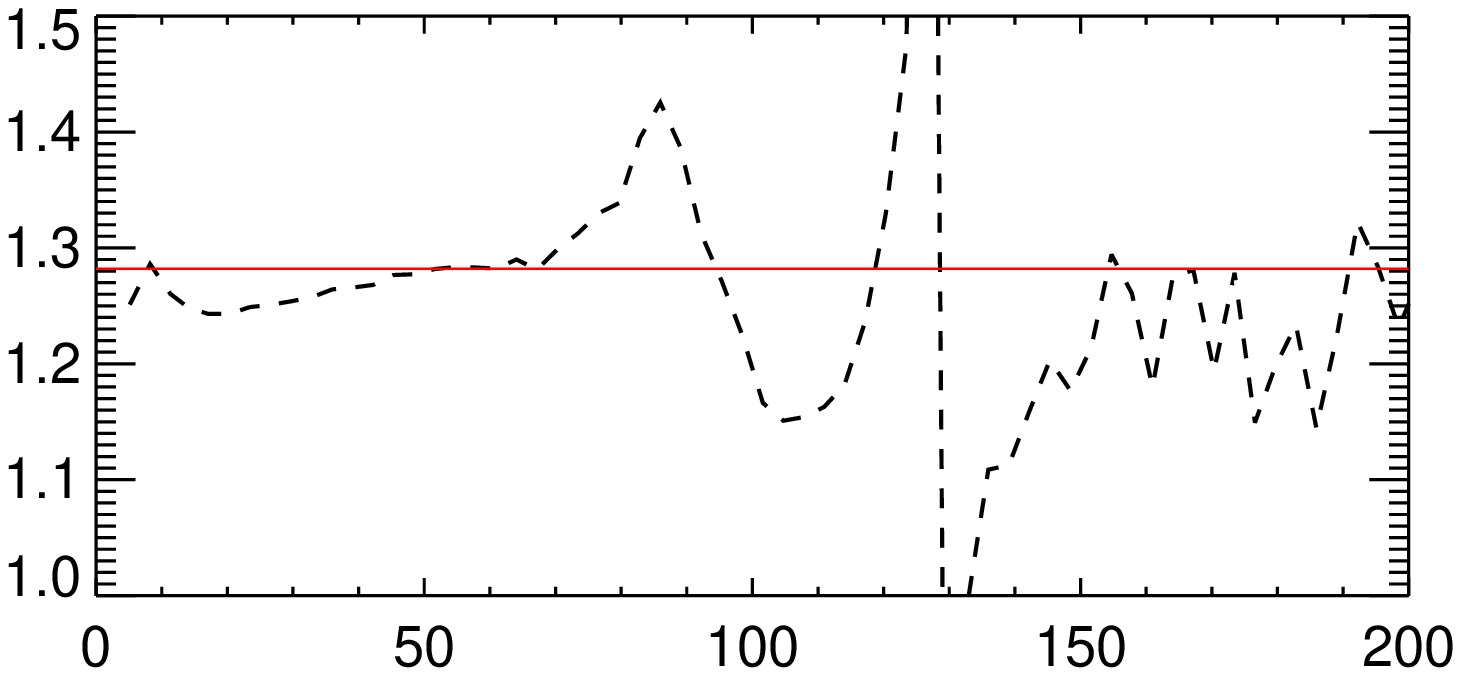}
\put(-132,30){\rotatebox[]{90}{$\xi^z(r)/\xi(r)$}}
\put(-100,20){\textsc{patchy}}
\put(-80,-5){$r$ [$h^{-1}$ Mpc]}
\end{minipage}
\begin{minipage}{0.2\textwidth}
\vspace{-4.5cm}
\caption{
\label{fig:corrquo}
Quotient between the averaged correlation functions in redshift- ($\xi^{z}(r)$) and real-space ($\xi(r)$) for the 8 sub-volumes of the BigMultiDark simulation (top) and for the 50  \textsc{patchy} realizations (bottom). The red line corresponds to the Kaiser factor prediction. 
}
\end{minipage}
\vspace{-.15cm}
\\
\hspace{-.1cm}
\includegraphics[width=4.6cm]{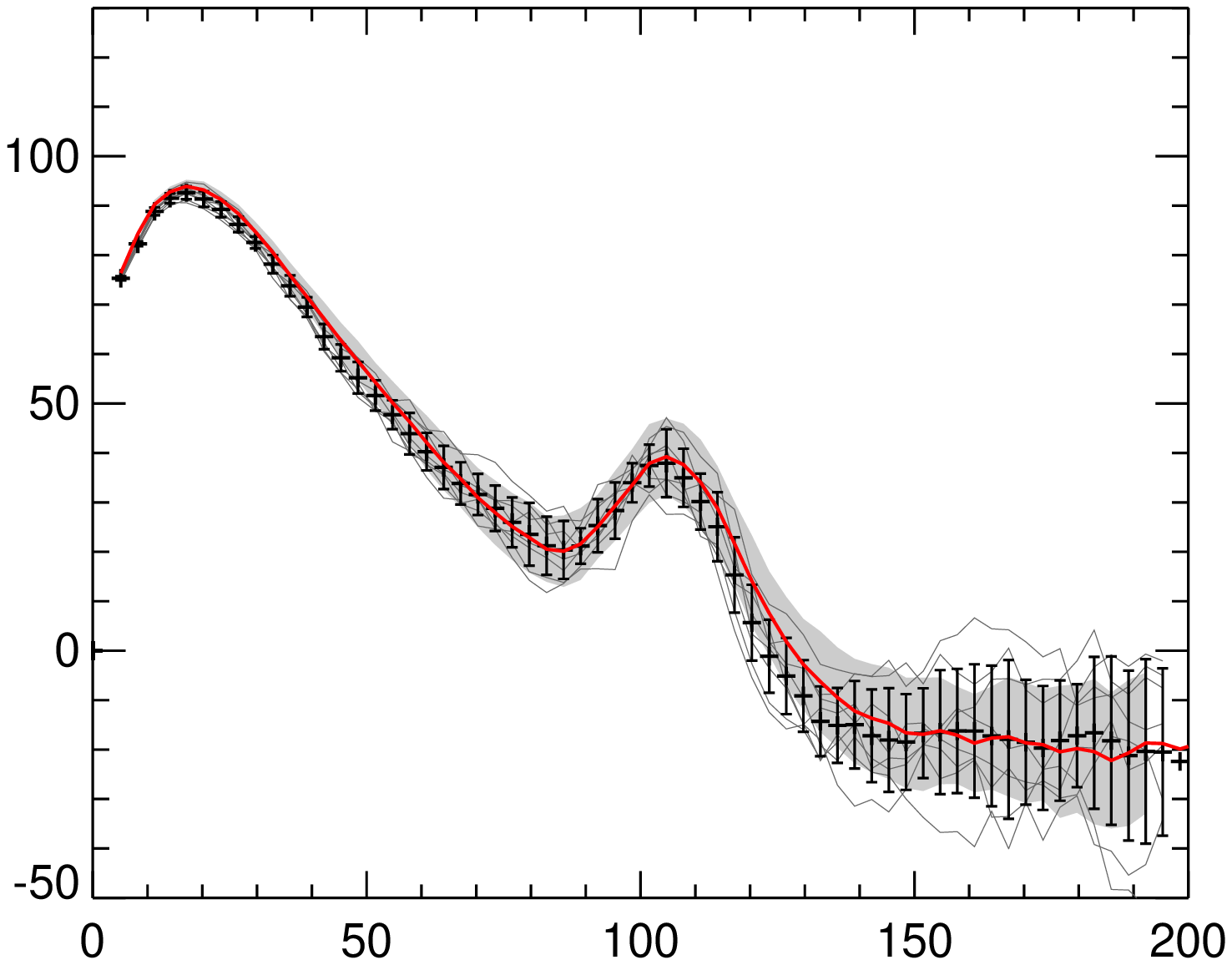}
\put(-70,90){real-space}
\put(-132,65){\rotatebox[]{90}{$\xi(r)\times r^{2}$}}
\put(-80,2){$r$ [$h^{-1}$ Mpc]}
\hspace{-0.835cm}
\includegraphics[width=4.6cm]{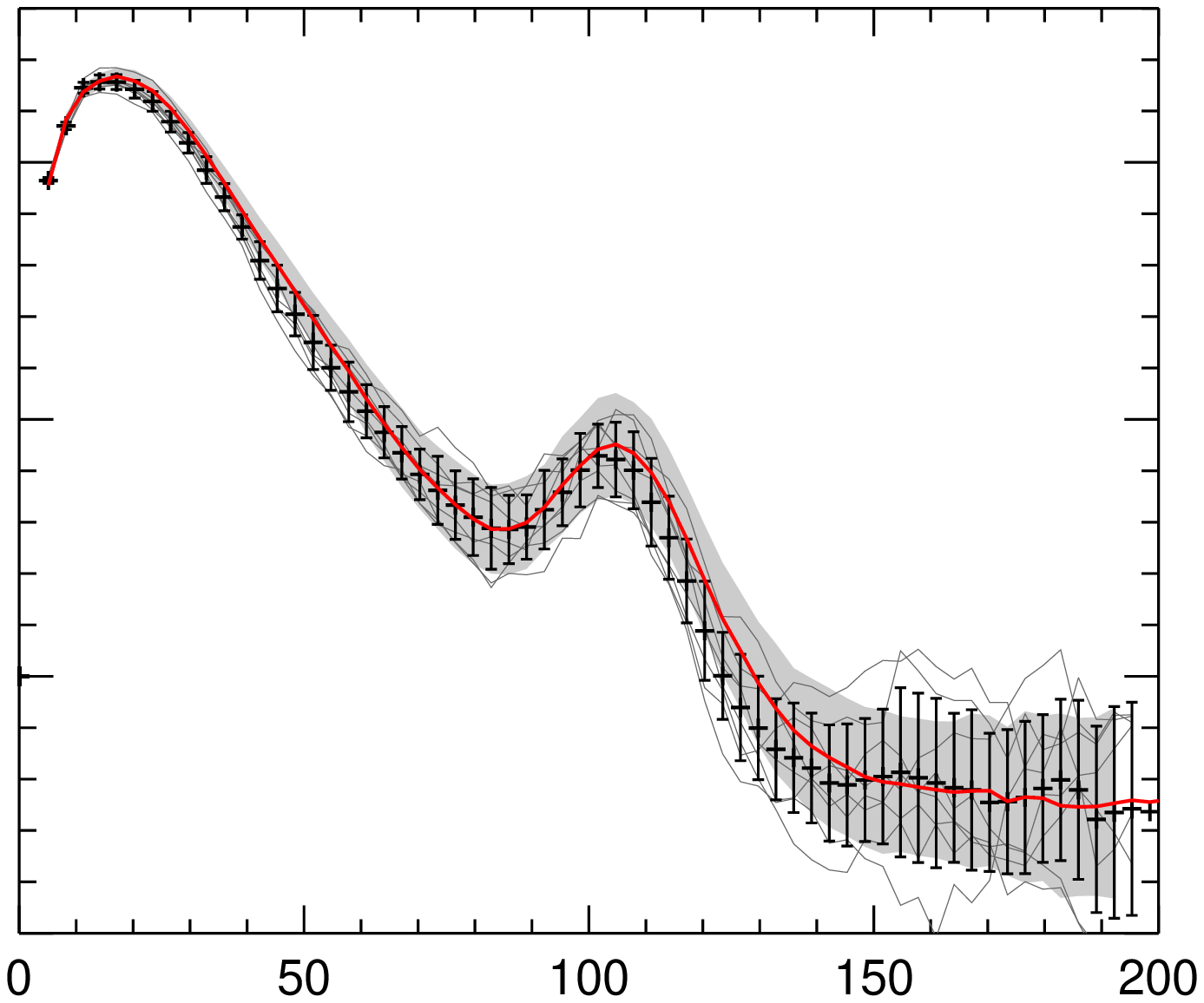}
\put(-70,90){redshift-space}
\put(-80,2){$r$ [$h^{-1}$ Mpc]}
\vspace{-.15cm}
\caption{
\label{fig:corr}
Correlation functions of the \textsc{patchy} simulations vs the BigMultiDark $N$-body simulation. The red line corresponds to the mean of the \textsc{patchy} realizations with the corresponding 1-sigma region in grey. Crosses: mean over the {BigMultiDark} sub-volumes. The error bars indicate 1-sigma regions for the $N$-body case. Each of the 8 sub-volumes drawn from the $N$-body simulation is represented by the grey lines. The left panel shows real-space, while the right panel shows redshift-space.
}
\end{figure}

We make a partition of the {BigMultiDark} box into 8 sub-volumes of equal size $(1250\,h^{-1}{\rm\,Mpc})^3$. This permits us to get rough estimates of the variance in the power spectra and correlation functions due to cosmic variance.
Since our approach yields number counts in cells on a mesh, we define the reference power spectrum  as the mean of the ones corresponding to the  halo overdensity field in the sub-volumes gridded with nearest-grid-point (NGP). We choose a mesh  of $512^3$  cells to reach a resolution of cell size $2.4\, h^{-1}$ Mpc.
We have explored the parameter space of our model \{$\alpha,\beta,\delta^{\rm low},\delta^{\rm high},g$\} running \textsc{patchy} with  $512^3$ particles in volumes of $(1250\,h^{-1}{\rm\,Mpc})^3$  to maximize the fit to the reference power spectrum in the relevant range for BAOs. This leads to an inconsistency with the large-scale modes which are included in BigMultiDark, but not in the \textsc{patchy} realizations. We will investigate the impact of this approximation in a future work. Our criterion for the parameter selection  is based on reaching better than 2\% accuracy in the range $0.07< k < 0.4\,h$ Mpc$^{-1}$. 
Using a set of parameters, which meet our criteria, we perform 50 random seeded realizations with \textsc{patchy}.
We have checked that the halo number density from the realizations is compatible with the expected number density  (about 70\% of the realizations  lie within the 1-sigma region of the $N$-body simulation with a very close mean).
A slice through the distribution of the dark matter and the corresponding halo sample are presented in Fig.~\ref{fig:2dplots}. The dark matter slice clearly shows the nonlinear cosmic web. Looking carefully, one can distinguish resolution effects in  voids. However, these should not affect our mocks as we do not consider massive halos in low density regions. This issue should be revisited when generating catalogues for low mass halos.
Fig.~\ref{fig:ps} shows the power spectra comparing the results between \textsc{patchy} and the BigMultiDark $N$-body simulation. On the left panel we find an agreement between the model and the reference power spectrum within 2\% up to $k\sim 1\,h$ Mpc$^{-1}$ in real-space. The same is shown in redshift-space on the right panel.  We have checked the \textsc{patchy} performance neglecting over-dispersion using the Poisson disstribution also in the high density regime. In this case the deviation from the reference power spectrum can be larger than 10\%, being in particular above 10\% at $k = 0.4\,h$ Mpc$^{-1}$ (see dotted line in the left-lower panel on Fig.~\ref{fig:ps}).
Fig.~\ref{fig:corrquo} shows that the Kaiser factor $K$ \citep[][]{Kaiser-87}\footnote{$K=1+\frac{2}{3}\beta+\frac{1}{5}\beta^2\sim1.28$, with $\beta=f/b$, $f$ being the growth rate, and $b\sim2$ being the linear bias for our mock LRG sample.}, the BAO damping, and the excess of power at scales lower than the BAO peak, are well reproduced with our redshift-space distortion model.
The correlation functions for both real- and redshift-space are shown in  Fig.~\ref{fig:corr}. We can see that n all bins PATCHY is compatible, within the error bars, with the BigMultiDark $N$-body  simulation down to scales $\lsim5\,h^{-1}$ Mpc.
The dispersion in the correlation functions from the $N$-body and the \textsc{patchy} realizations are remarkably similar.

\section{Conclusions and Discussion}

\label{sec:conc}

In this work we have combined structure formation perturbation theory with a local, nonlinear, stochastic bias including redshift-space distortions to generate mock halo catalogues  of a given number density using the novel \textsc{patchy}-code.  
We have also shown the importance of treating over-dispersion in high density regions to reach precisions of about 2\% in the power spectrum.
The advantage of our approach is manifold. First, we use an improved efficient one-step solver (Augmented Lagrangian Perturbation Theory: ALPT) yielding a smooth density field on a mesh. Second, we use a statistical model applied to that field to produce  a distribution of halos with power spectra matching those obtained from the BigMultiDark $N$-body simulations, as opposed to other methods, which rely on the halos obtained with the corresponding approximation (e.g. \textsc{PThalos}, \textsc{Pinocchio} or \textsc{Cola}). In this way, we circumvent two problems: the inaccuracy of the perturbative approach and the large number of particles required  to resolve the halo population of interest within volumes. 
In our tests we have used $\sim50$ times less particles than for the BigMultiDark  simulation in the corresponding volume.
Finally, we have modeled redshift-space distortions, reproducing the correlation function of $N$-body simulations with the same level of accuracy as in real-space.

Still a number of issues have to be investigated, such as the  performance of the method as a function of redshift (including light-cones), cosmological parameters, nonlocal bias, and different halo number densities. Higher order statistics (skewness, kurtosis, three-point correlation functions or bispectrum) are expected to be reasonably well  modeled as we are using improved versions of 2LPT correcting for the damped power spectrum. We will investigate all these issues in detail for mass production of mock catalogues and plan to make \textsc{patchy} publicly available. 

In summary, \textsc{patchy} proves to  be  an especially efficient and accurate  method based on perturbation theory to generate mock catalogues including nonlinear, scale-dependent and stochastic biasing with redshift-space distortions, which can be used to compute covariance matrices  for large galaxy surveys.

\vspace{0.1cm}
\begin{flushleft}
{\bf Acknowledgments}
\end{flushleft}
FSK thanks V.~M\"uller for encouraging discussions. Special thanks to A.~Klypin, S.~He{\ss}, S.~Gottl{\"o}ber and C.~Scoccola. We thank R.~E.~Angulo and C.~H.~Chuang for comments on the manuscript.  The work was initiated under the HPC-Europa2 project (project number: 228398) with the support of the European Commission - Capacities Area - Research Infrastructures.  The MultiDark Database used in this paper and the web application providing online access to it were constructed as part of the activities of the German Astrophysical Virtual Observatory as result of a collaboration between the Leibniz-Institute for Astrophysics Potsdam (AIP) and the Spanish MultiDark Consolider Project CSD2009-00064.  The  BigMD simulation suite have been performed in the Supermuc supercomputer at LRZ using time granted by PRACE.
GY and FP acknowledge support from the Spanish MINECO under research grants  AYA2012-31101, FPA2012-34694, AYA2010-21231, Consolider Ingenio SyeC CSD2007-0050 and  from Comunidad de Madrid under  ASTROMADRID  project (S2009/ESP-1496). 
\vspace{-.6cm}